\documentclass[final,3p,times,authoryear]{elsarticle}

\usepackage{amssymb}
\usepackage{amsmath}
\usepackage{float}
\usepackage{graphicx}
\usepackage{xcolor}
\usepackage{siunitx}
\usepackage{soul}

\usepackage{lineno}
\usepackage{url}
\newcommand{\vect}[1]{\mathbf{#1}}        
\newcommand{\mat}[1]{\mathbf{#1}}         
\newcommand{\tens}[1]{\mathcal{#1}}       
\newcommand{\uvec}[1]{\vec{\mathbf{#1}}}  
\newcommand{\nvec}[1]{\hat{\mathbf{#1}}}  

\journal{Journal TBD}

\begin{document}

\begin{frontmatter}



\title{Automated Burgers Vector Identification for Individual Dislocations in Bulk Crystals}

\author[1,2]{Abderrahmane Benhadjira\fnref{fn1}\corref{cor2}}
\author[1]{Carsten Detlefs}
\author[1,2]{Vincent Favre-Nicolin}
\author[3]{Henning Friis Poulsen}
\author[4]{Grethe Winther}
\author[1]{Can Yildirim}
\author[3]{Sina Borgi\fnref{fn1}\corref{cor1}}

\cortext[cor1]{Corresponding author e-mail: borgi@dtu.dk}
\cortext[cor2]{Contributing author e-mail: abderrahmane.benhadjira@esrf.fr}
\fntext[fn1]{These authors contributed equally to the presented work.}

\affiliation[1]{organization={European Synchrotron Radiation Facility}, 
          city={Grenoble}, 
          postcode={38043}, 
          country={France}}
          
\affiliation[2]{organization={Univ. Grenoble Alpes}, 
          city={Grenoble}, 
          postcode={38400}, 
          country={France}}

\affiliation[3]{organization={Department of Physics, Technical University of Denmark}, 
          city={Kgs. Lyngby}, 
          postcode={2800}, 
          country={Denmark}}

\affiliation[4]{organization={Department of Mechanical Engineering, Technical University of Denmark}, 
          city={Kgs. Lyngby}, 
          postcode={2800}, 
          country={Denmark}}

\begin{abstract}
Weak-beam imaging in dark-field X-ray microscopy (DFXM) can resolve individual dislocations in bulk crystals, but assigning Burgers vectors from the resulting contrast typically requires manual comparison with forward simulations. Here, we train a physics-informed convolutional neural network (CNN) on geometrical optics simulations of isolated dislocations in face-centred cubic (FCC) aluminium, incorporating crystallographic constraints into the learning process. The model identifies Burgers vectors from weak-beam integrated rocking-curve images. On synthetic test data, the model achieves an accuracy of approximately 93\%. In an experimental cross-slip case, the constrained model assigns 72.7\% of the layer-wise predictions to the reference Burgers vector. These results show that simulation-trained, physics-informed CNNs represent a step toward automated dislocation identification in DFXM.
\end{abstract}




\end{frontmatter}

\section{Introduction}\label{sec:Introduction}
The plastic deformation of crystalline materials is governed by the nucleation, motion, interaction, and accumulation of line defects known as dislocations, which accommodate permanent shape changes at the atomic scale \citep{Hirth1992}. These dislocations form complex networks that evolve under stress and strongly influence key mechanical properties such as strength, ductility, and toughness. As their density and arrangement change during deformation, they promote the formation of hierarchical structures such as planar dislocation networks delineating more equiaxed cells\citep{Winther2007,MOUSSA2017} or dislocation interaction with grain boundaries \citep{Guo_2020}, ultimately shaping the macroscopic behavior of metals and other crystalline solids. Beyond mechanical response, dislocations are also important in semiconductor materials, where their strain fields and defect cores can introduce electronic states, affect carrier transport, and act as radiative or non-radiative recombination centers. Consequently, they can influence optical and electronic properties such as photoluminescence, carrier lifetime, leakage current, and device efficiency \citep{Sauer1985,Schroter1995,Speck1999,Reiche2016}. Capturing dislocation behavior is therefore central to both fundamental materials science and engineering applications, including the design of structural materials, microelectronic components, optoelectronic devices, and energy technologies.

To model these processes, multiscale frameworks have been developed ranging from discrete dislocation dynamics (DDD) simulations \citep{Devincre2011,Po2014} to continuum dislocation dynamics (CDD) simulations \citep{Mohamed2015,PACHAURY2022104861}. However, these models heavily depend on having realistic initial spatial dislocation distributions as well as statistics on dislocation interactions and mobilities at the microscale. Accurate, spatially resolved experimental data on individual dislocations in real materials is crucial for validating and refining these simulations. Despite decades of progress in electron microscopy and diffraction methods, a significant need remains for experimental techniques that can non-destructively resolve individual dislocations in three dimensions (3D) within bulk samples under realistic conditions \citep{Lang1997,Ludwig2001,Simons2015,Brennan2022,Hanschke2012DiffractionLaminography,Danilewsky2020NicePictures,Tanuma2012Microbeam3DSiC}.

The characterization of dislocations has previously relied heavily on electron microscopy techniques. Transmission electron microscopy (TEM) remains the gold standard for imaging individual dislocations with near-atomic spatial resolution and has played a central role in developing our understanding of dislocation core structures and interactions \citep{Williams2009}. More recently, electron backscatter diffraction (EBSD), and in particular its high-resolution variant (HR-EBSD), has demonstrated the ability to map dislocation-induced lattice rotations and even resolve individual dislocations under favorable conditions \citep{MOUSSA2017,Ernould2022}. However, these techniques are inherently limited to two-dimensional (2D) surfaces or thin-foil samples, making them less suited for studying the dynamics of evolving 3D dislocation networks under stress in bulk specimens.

To overcome these limitations, dark-field X-ray microscopy (DFXM) has emerged as a powerful alternative for non-destructive imaging of dislocations deeply embedded within crystalline materials. DFXM uses high-energy synchrotron X-rays and objective-based imaging in Bragg diffraction geometry to probe specific crystallographic planes at depth, offering sub-micrometer spatial resolution and sensitivity to strain fields on the order of $10^{-4}$ radians \citep{Simons2015b,Jakobsen2019,Yildirim2023}. By tuning to weak-beam conditions, DFXM selectively enhances contrast from regions of lattice distortion, allowing individual dislocations to be imaged and tracked \textit{in situ} during mechanical loading \citep{Dresselhaus-Marais2021,Frankus2025}. Despite these advantages, interpreting weak-beam images still requires significant manual input or comparison to simulations, and the volume of data acquired in 3D or dynamic experiments can quickly become a bottleneck for analysis. As such, there is a growing need for automated, scalable approaches that can extract dislocation-level information from DFXM data with minimal user intervention.

Recent advances in DFXM have demonstrated the technique's potential for resolving not only static dislocation structures, but also their dynamics and interactions in 3D. For example, \cite{Dresselhaus2021} and \cite{Yildirim2023} imaged individual dislocations in bulk aluminium with sub-micrometer resolution, while \cite{Frankus2025} captured the 3D evolution of dislocation pile-ups during tensile loading, offering experimental insight into collective dislocation behavior under stress. In addition to these efforts, \cite{Pal2025} extended the classical invisibility criterion from TEM to DFXM by analysing asymmetries in rocking curves across multiple tilt directions, providing a framework for deducing Burgers vectors from angular contrast patterns.

Building on some of these developments, the work in \cite{Borgi2025} introduced a framework that uses geometrical optics (GO) to simulate weak-beam DFXM images for isolated dislocations in face-centred cubic (FCC) aluminium crystals \citep{Poulsen2017,Poulsen2021,Borgi2024}. By generating libraries of simulated images that span all possible combinations of slip planes, Burgers vectors, and line directions, it was demonstrated that the identity of a dislocation can be inferred from a single weak-beam image through cross-correlation, provided that the dislocation is well isolated and the imaging conditions are optimized. This established a new paradigm for direct, quantitative dislocation identification from sparse 2D data, offering a valuable complement to more time-consuming 3D reconstructions. 

Although identification via cross-correlation with forward-modelled images can be accurate, it requires comparisons across many possible configurations and may not scale easily to high-throughput or real-time workflows. In time-resolved experiments, where dislocation structures evolve under applied stress, the amount of image data increases, placing further demands on analysis pipelines. Moreover, as experimental setups diversify, e.g.~probing different reflections, geometries, and materials, the parameter space of possible dislocation images also expands. These trends highlight the need for automated methods that can generalize across experimental conditions, reduce user input, and provide fast and reliable dislocation identification at scale.

Machine learning (ML) has been increasingly adopted in microscopy and diffraction to accelerate data interpretation and enable real-time feedback during experiments. In electron microscopy, ML has been used to improve orientation mapping accuracy and classify deformation mechanisms from diffraction-based image data such as EBSD and scanning transmission electron microscopy (STEM) \citep{DING2021,Wittwer2022}. For instance, \cite{Zhang2019} employed convolutional neural networks (CNNs) to extract dislocation microstructures from EBSD data. Similarly, \cite{Yassar2010} used artificial neural networks (ANNs) to predict the flow stress of micropillars, incorporating dislocation density and model size as input features. \cite{Tao2024} used ML models to predict micropillar compression behavior based on DDD simulations. Additionally, \cite{Hiemer2023} applied kernel ridge regression (KRR) to establish links between plasticity and dislocation characteristics, analysing the influence of strain rate, dislocation density, and strengthening mechanisms in FCC metals. In X-ray diffraction applications, \cite{Masto2024} used a U-net for inpainting Bragg coherent diffraction patterns affected by detector gaps, and \cite{Yanxon2023} used traditional ML models to classify artifacts in X-ray diffraction data. These advances demonstrate the potential of supervised learning for extracting physically meaningful information from high-dimensional image data, and suggest that similar strategies could enhance the scalability and automation of dislocation analysis in DFXM.

A second, pragmatic motivation is acquisition speed: learning-based analyses can sometimes tolerate lower signal-to-noise ratios (SNR), allowing a trade-off between dose and throughput. In EBSD, physics-guided dictionary indexing and related methods routinely index patterns at reduced dwell times and are explicitly robust to noise, enabling near-real-time throughput \citep{DeGraef_2020,Lenthe2019}. Recent work pushes this further by denoising or reconstructing patterns from subsampled scans with ML, sustaining indexing fidelity at lower SNR \citep{Krishna2023,Broad2025}. By the same logic, well-calibrated DFXM setups may allow shorter exposures while retaining reliable predictions, increasing acquisition rates.

In this work, we present a supervised ML approach for identifying the Burgers vector of an isolated dislocation in weak-beam integrated (WBI) rocking-curve images \cite{Benhadjira2026}. CNNs based on the ResNet-18 architecture \cite{resnet} are trained on GO simulations to associate image features with specific Burgers vectors across a range of dislocation types and orientations. We compare a purely data-driven model with a crystallographically constrained model, evaluate uncertainty estimates, test transfer between reflections, and apply the trained models to experimental DFXM images of isolated dislocations. Once trained, the model assigns a Burgers-vector label to new cropped images within milliseconds, reducing the need for manual comparison against simulated libraries \citep{Borgi2025}. Focusing on isolated dislocations in WBI rocking curves, we assess the potential and present limitations of simulation-trained CNNs for scalable Burgers vector identification in bulk materials.

\begin{figure}
    \centering
    \includegraphics[width=0.9\linewidth]{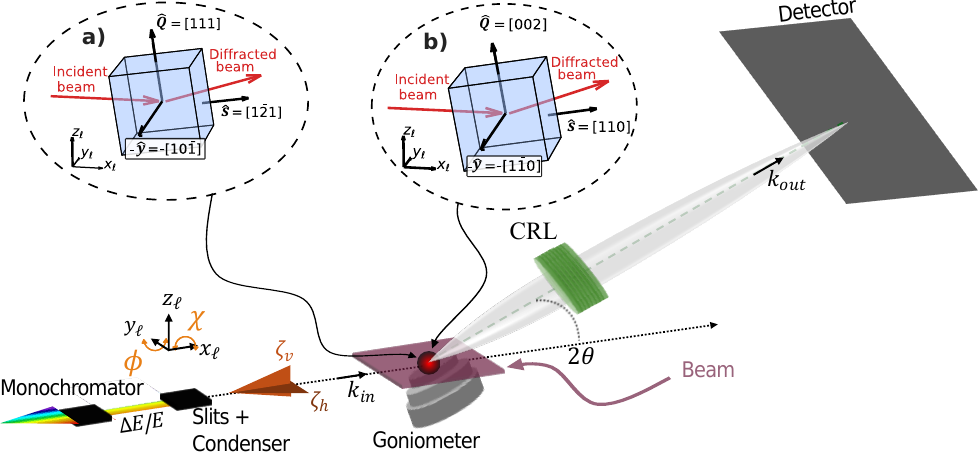}
    \caption{Schematic of the DFXM geometry used for the simulations and experiments. A monochromator defines the relative energy bandwidth $\Delta E/E$, while slits and a one-dimensional condenser set the beam size, incoming divergences $(\zeta_v,\zeta_h)$, and illuminated sheet height $\Delta z_\ell$. The diffracted beam at angle $2\theta$ is collected by the compound refractive lens (CRL) and recorded on the downstream detector. Insets (a) and (b) show the crystal orientations for the two reflections used in this work, with diffraction vector $\nvec{Q}$, the crystallographic direction $\nvec{y}$ parallel to $y_\ell$, and surface-normal direction of the sample, $\nvec{s}$ in the diffracted-beam plane. The $\nvec{y}$ direction is plotted inversely for visualization. The $\phi$ scan rocks the sample about $y_\ell$, while the $\chi$ scan rolls the sample about $x_\ell$.}
    \label{fig:DFXM}
\end{figure}

\begin{figure}
    \centering
    \includegraphics[width=0.7\linewidth]{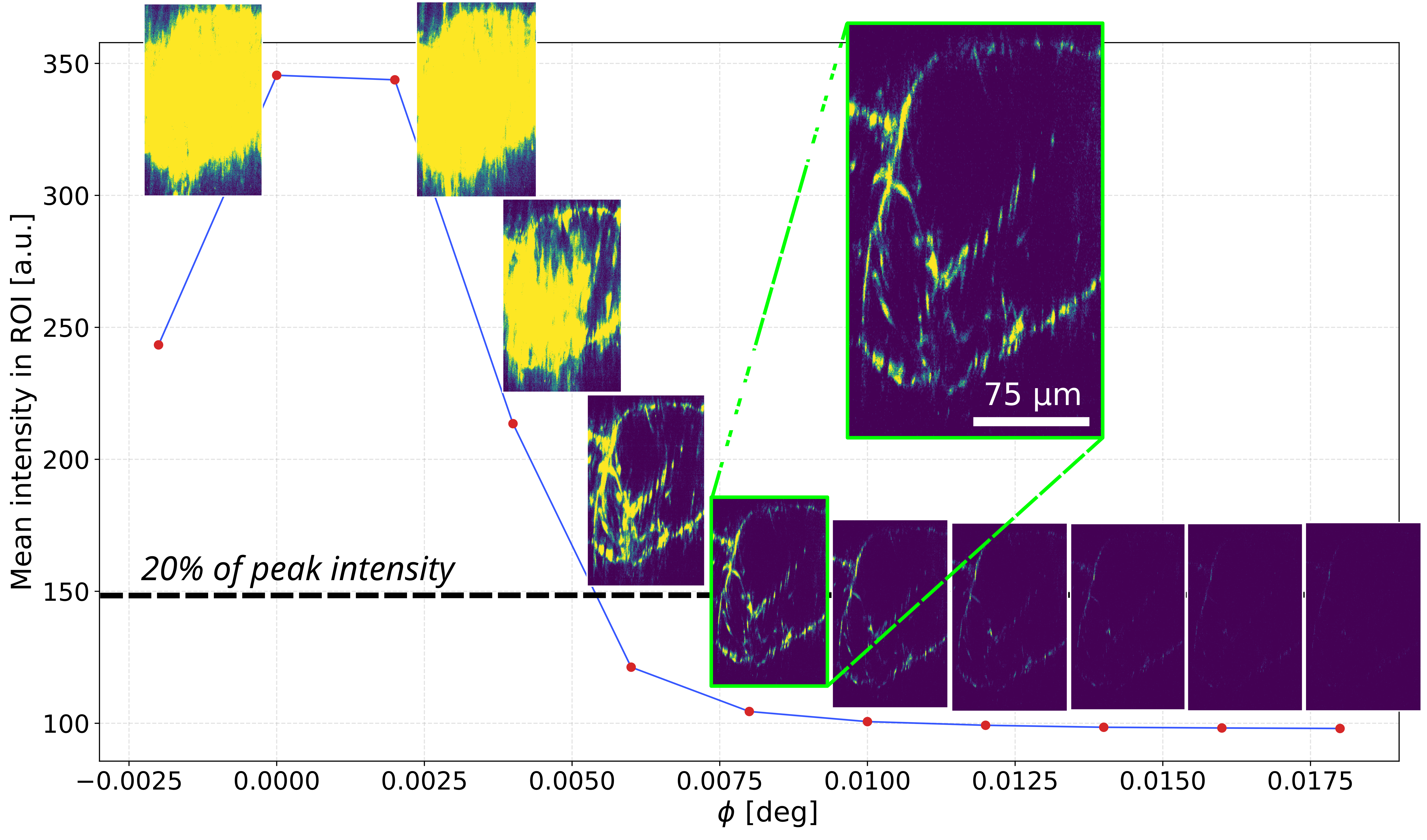}
    \caption{
Rocking curve measured around the nominal Bragg condition, $\phi = 0^\circ$, for the
$\mathbf{Q} = [111]$ reflection of the aluminium single crystal.
The blue curve shows the mean intensity, in the ROI, of the acquired diffraction images as a
function of sample rotation. Red markers highlight the data points along the rocking curve and raw images are added as insets at each point along the curve. The insets have the same intensity ranges for their color range, to highlight the effect of the strong-beam condition and weak-beam conditions effect on dislocation contrast in DFXM, with all the images having a mean intensity above the dashed black line being in the so-called strong-beam condition. There is a zoom in of the inset at $\phi=0.008^\circ$, here one can see a domain in the ROI where multiple dislocations have strong contrast to the background noise. 
}

    \label{fig:wbi}
\end{figure}

\section{Methods}\label{sec:Methods}
The training datasets were generated from geometrical-optics (GO) DFXM simulations, following \cite{Borgi2025}. A schematic of the experimental geometry is shown in Fig.~\ref{fig:DFXM}. The simulated dislocation configurations were specified by the Burgers vector, line direction, and slip plane, as described in Sec.~\ref{sec:digital_phantoms}. For each configuration, two diffraction vectors were considered: $\uvec{q}_{hkl}=[002]$ and $\uvec{q}_{hkl}=[\bar{1}\bar{1}\bar{1}]$.

The model was trained using simulated weak-beam DFXM images obtained within the GO approximation. In a rocking curve, the weak-beam regime corresponds to the tails away from the nominal Bragg condition, as illustrated in Fig.~\ref{fig:wbi}. There the strong-beam condition, where the crystal lattice is in diffraction and the image intensity is high, the contrast for dislocations is low. However, when the sample is rotated away from the nominal condition, the mean intensity decreases and the diffraction contrast changes, as highlighted in Fig.~\ref{fig:wbi} with the large inset at $\phi = 0.008^\circ$. In the weak-beam condition the dislocations have strong contrast to the background noise and can be seen as blobs or streaks depending on the dislocation character. In this work, WBI images were obtained by summing all images acquired under the weak-beam conditions along the rocking curve, only excluding the strong-beam condition images with intensities $\geq20\%$ of the peak mean intensity, marked with the dashed line in Fig.~\ref{fig:wbi}. In this regime, the diffracted intensity is reduced and the contrast is dominated by strain-induced deviations from the Bragg condition, making dislocation distortion fields more visible. By contrast, images near the rocking-curve maximum correspond to the strong-beam condition, where multiple scattering and dynamical diffraction effects can dominate the image contrast. Because these effects are not captured by the GO model, strong-beam images were excluded from training.

\subsection{Ground truth: Geometrical Optics}\label{subsec:Data}

In DFXM, forward modelling of dislocation contrast can be efficiently achieved using a GO approach. In the weak-beam regime, where dynamical diffraction effects are minimal, GO approximates diffraction via ray tracing and couples the local deformation fields of the crystal to the anisotropic angular and energy acceptance of the microscope. This provides a computationally inexpensive framework that enables experiment planning, real-time interpretation during beamtime, and post-experimental analysis of dislocation structures.  

The GO formalism is divided into reciprocal- and direct-space stages and follows \cite{Poulsen2021}. In reciprocal space, we sample ray directions uniformly over the objective's acceptance cone, the numerical aperture (NA), and their transmission is then weighted by the objective's angular acceptance. We parametrize directions by the out-of-plane azimuth angle, $\eta$, and the in-plane deviation from the nominal scattering angle, $\Delta 2\theta = 2\theta -2\theta_0$, (cf. Fig.~\ref{fig:DFXM}). The probability that a ray with direction $(\eta, 2\theta)$ reaches the detector is governed by the objective's angular acceptance. Monte Carlo sampling of this angular domain, together with the incident energy spread, $\varepsilon = \Delta E/E$, and small vertical/horizontal beam divergences, $(\zeta_v, \zeta_h)$, is mapped to a local reciprocal-space frame, $(q_{\mathrm{rock}}, q_{\mathrm{roll}}, q_{\parallel})$, in crystal coordinates. In the small-angle limit, the components are
\begin{align}
\vect{q}_{\mathrm{rock}} &= -\frac{\zeta_v}{2} - \frac{\Delta 2\theta}{2}, \\
\vect{q}_{\mathrm{roll}} &= -\frac{\zeta_h}{2\sin\theta} - \cos \theta ~\eta, \\
\vect{q}_{\parallel} &= \varepsilon + \frac{1}{\tan \theta}\left(-\frac{\zeta_v}{2} + \frac{\Delta 2\theta}{2}\right),
\end{align}
The imaging-frame components $(\vect{q}'_{\mathrm{rock}}, \vect{q}_{2\theta})$ follow by a rotation through the Bragg angle, $\theta$, within the diffraction plane,
\begin{align}
\vect{q}'_{\mathrm{rock}} &= \cos\theta \, \vect{q}_{\mathrm{rock}} + \sin\theta \, \vect{q}_{\parallel}, \\
\vect{q}_{2\theta} &= -\sin\theta \, \vect{q}_{\mathrm{rock}} + \cos\theta \, \vect{q}_{\parallel}.
\end{align}
Voxelization of $(\vect{q}'_{\mathrm{rock}}, \vect{q}_{\mathrm{roll}}, \vect{q}_{2\theta})$ over the sampled rays yields the three-dimensional resolution kernel $\mathrm{Res}_{qi}$, which encodes the instrument's angular/energy acceptance. Throughout, we use normalized diffraction vectors, $\vect{q}=\Delta\vect{Q}/|\vect{Q_0}|$, appropriate to the narrow range in reciprocal space probed in DFXM.

In direct space, this kernel is coupled to the elastic fields of dislocations via the deformation gradient tensor, $\tens{F}$. For a dislocation with Burgers vector \(\nvec{b}\), slip-plane normal \(\nvec{n}\), and line direction \(\nvec{t}\), $\tens{F}$ is written as a superposition of edge and screw ($\tens{F}_{edge}, \tens{F}_{screw}$) components:
\begin{equation}
\tens{F} = \tens{F}_{\mathrm{screw}}\cos\alpha + \tens{F}_{\mathrm{edge}}\sin\alpha,
 \end{equation}
where \(\alpha\) is the angle between \(\nvec{b}\) and \(\nvec{t}\). In grain coordinates, to easily fit to the corresponding reciprocal space components, the micro-mechanical model is described as
\begin{equation}
\tens{H}_g = (\tens{F}_g)^{-T} - \mat{I},
\end{equation}
with \(\tens{F}_g\) the deformation gradient in the grain frame and \(\mat{I}\) the identity matrix. The scattering vectors that determine the simulated image are then constructed from the scattering vector in sample-, grain-, and imaging-space respectively

\begin{align}
\vect{q}_s &= \mat{U}_s \tens{H}_g \uvec{q}_{hkl}, 
\end{align}

where \(\mat{U}_s\) transforms from sample to grain coordinates and $\uvec{q}_{hkl}$ is the normalized scattering vector along the Miller indices $(hkl)$.

\begin{align}
\vect{q}_g &= \vect{q}_s +
\begin{pmatrix}
\phi - \Delta \theta \\
\chi \\
\Delta \theta / \tan \theta
\end{pmatrix},
\end{align}

where $\phi$, $\chi$, and $\Delta\theta$ describe small offsets in the goniometer degrees of freedom. 

\begin{align}
\vect{q}_i &= \mat{\Theta} \vect{q}_g,
\end{align}

and \(\mat{\Theta}\) applies the Bragg-angle rotation to imaging space. By accumulating contributions from all illuminated voxels through \(\text{Res}_{qi}\), a synthetic DFXM image is generated.  

This forward-modelling framework captures the characteristic weak-beam contrast of dislocations in DFXM. Its computational efficiency and interpretability make it suitable for generating the large simulated datasets used to train the ML model.

\begin{figure}[t]
    \centering
    \includegraphics[width=0.7\linewidth]{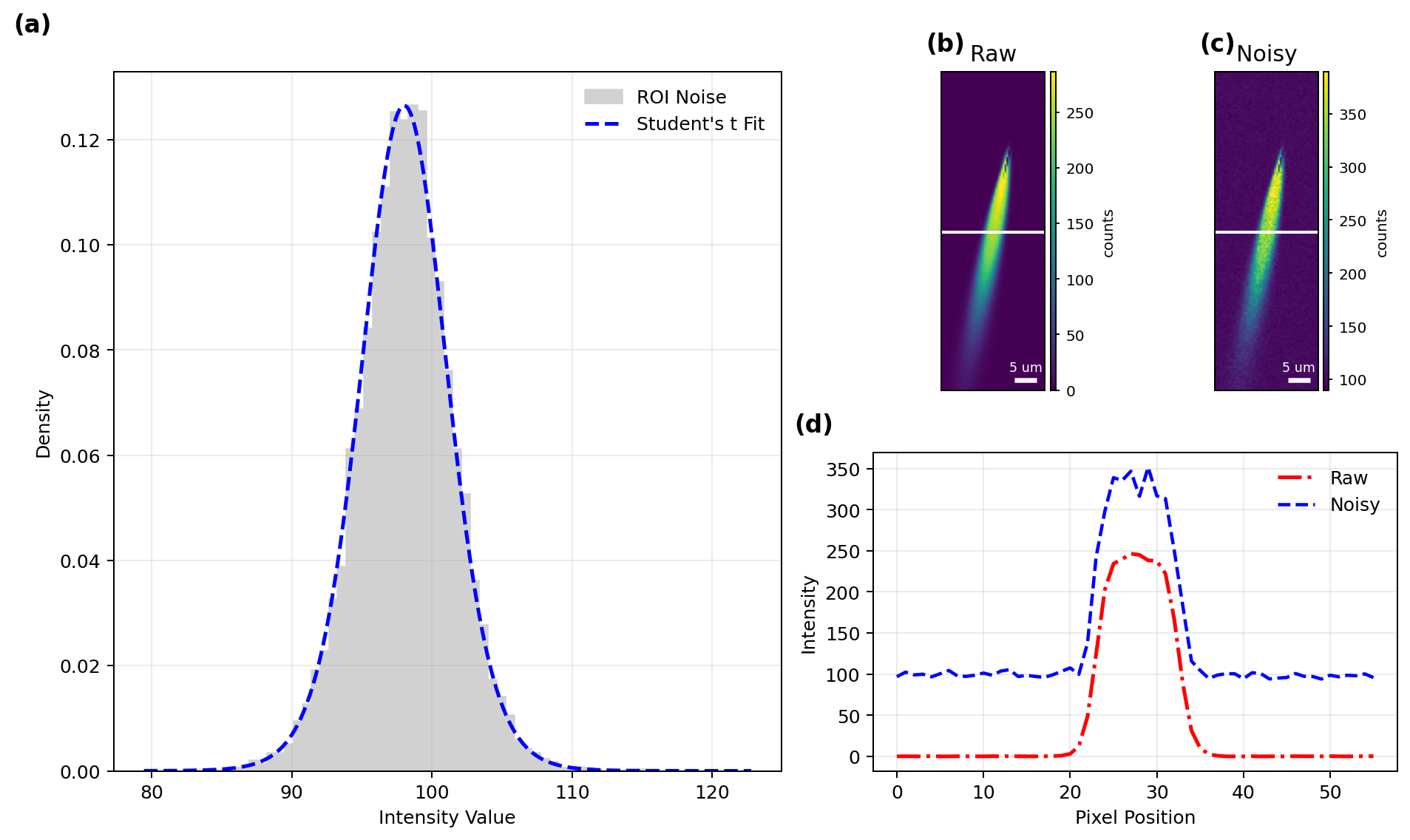}
    \caption{Noise model used to make simulated images closer to the experimental detector response. (a) Detector thermal background noise, shown in grey, fitted with a non-central Student's $t$ distribution, shown in blue, with $\nu = 13.067$, $\mu = 98.012$, and $\rho = 3.095$. (b) Raw simulated image containing an isolated dislocation near the centre of the field of view. (c) Same image after adding thermal and shot noise. (d) Line-intensity profiles across the raw and noisy images in (b) and (c).}
    \label{fig:snr}
\end{figure}

To emulate experimental photon statistics and strengthen the training, simple Poisson distributed shot noise was applied to each simulated image and then combined with a thermal noise component representing the detector background. The shot noise was applied to the signal intensity in each pixel of the simulated image, while the thermal noise was fitted to a dark frame (background signal) from an experimental measurement performed at the ID06 beamline at the European Synchrotron Radiation Facility (ESRF) \citep{Kutsal2019}. The best fit for the thermal noise was found to be a non-central Student's $t$-distribution with degrees-of-freedom $\nu = 13.067$, offset from zero (mean) $\mu = 98.012$, and a scale parameter $\rho = 3.095$, as shown in Fig.~\ref{fig:snr}a. The resulting noise model was then sampled pixel-wise and added to the simulated images, producing noisy counterparts that exhibit both the elevated background level and stochastic fluctuations observed experimentally. Representative examples of a raw and noisy image are shown in Fig.~\ref{fig:snr}b and c, respectively, with the corresponding line profiles in Fig.~\ref{fig:snr}d illustrating the effect of the noise on the signal while preserving the dislocation contrast.

To obtain the WBI images, a full rocking curve is simulated for each digital phantom described in Section~\ref{sec:digital_phantoms}. The rocking curve is generated by rotating the sample with the $\phi$ motor over $\pm 5 \times 10^{-4}$ degrees using 40 steps. The stack is then integrated over the weak-beam regime, where the intensity is below 20\% of the maximum intensity, excluding the strong-beam condition. All images are background subtracted, normalized, and stored with their Burgers vector labels for supervised CNN training.

\subsection{Digital phantoms}\label{sec:digital_phantoms}

To create each digital phantom, we use three right-handed coordinate frames, as described in \cite{Borgi2024} and \cite{Borgi2025}. In short, the \emph{laboratory} frame (subscript $\ell$) has $x_{\ell}$ parallel to the incident beam direction; $z_{\ell}$ is perpendicular to $x_\ell$; and $y_{\ell}$ completes the horizontal direction. The \emph{sample} frame (subscript $s$) is obtained by rotating about $y_{\ell}$ by the Bragg angle $\theta$. The \emph{grain} frame (subscript $g$) is related to the sample frame by the orientation matrix $\mat{U}$:

\begin{align}
    \mat{U} = 
    \begin{pmatrix}
    \nvec{s}_1 & \nvec{s}_2 & \nvec{s}_3 \\
    \nvec{y}_1 & \nvec{y}_2 & \nvec{y}_3 \\
    \nvec{Q}_1 & \nvec{Q}_2 & \nvec{Q}_3
    \end{pmatrix}.
\end{align}

where $\nvec{s}$ is the sample surface normal of the diffracted beam, $\nvec{y}$ is the grain direction parallel to the $y_\ell$ axis, and $\nvec{Q}$ is the diffraction vector. For the two reflections used in this work, the relation between the grain and laboratory coordinates is illustrated in Fig.~\ref{fig:DFXM} insets (a) and (b).

Similar to \cite{Borgi2025}, where a library of GO-based DFXM simulations was used to identify the Burgers vectors of isolated dislocations by cross-correlation with experimental weak-beam images, the present work uses the same simulation framework to generate a labelled training dataset for supervised learning. Each configuration corresponds to a single, straight dislocation embedded in an otherwise perfect FCC aluminium crystal and is defined by its slip plane, Burgers vector, and line direction. For FCC aluminium, the slip systems are of the type $\{111\}\langle 110\rangle$, with perfect-dislocation Burgers vectors $\nvec{b}=\frac{a}{2}\langle 110\rangle$. Treating $\nvec{b}$ and $-\nvec{b}$ as equivalent gives six unique $\langle 110\rangle$ Burgers-vector labels. These were combined with the four active $\{111\}$ slip planes, and the dislocation line direction was rotated in $\ang{10}$ increments, giving $6 \times 36 \times 4 = 864$ possible configurations. For each diffraction vector, 840 configurations were retained after removing 24 configurations that are invisible under the corresponding diffraction condition. This invisibility follows from the relation between the diffraction vector and the Burgers vector $\nvec{Q}\cdot\nvec{b}=0$.

\begin{figure}[t]
    \centering
    \includegraphics[width=\linewidth]{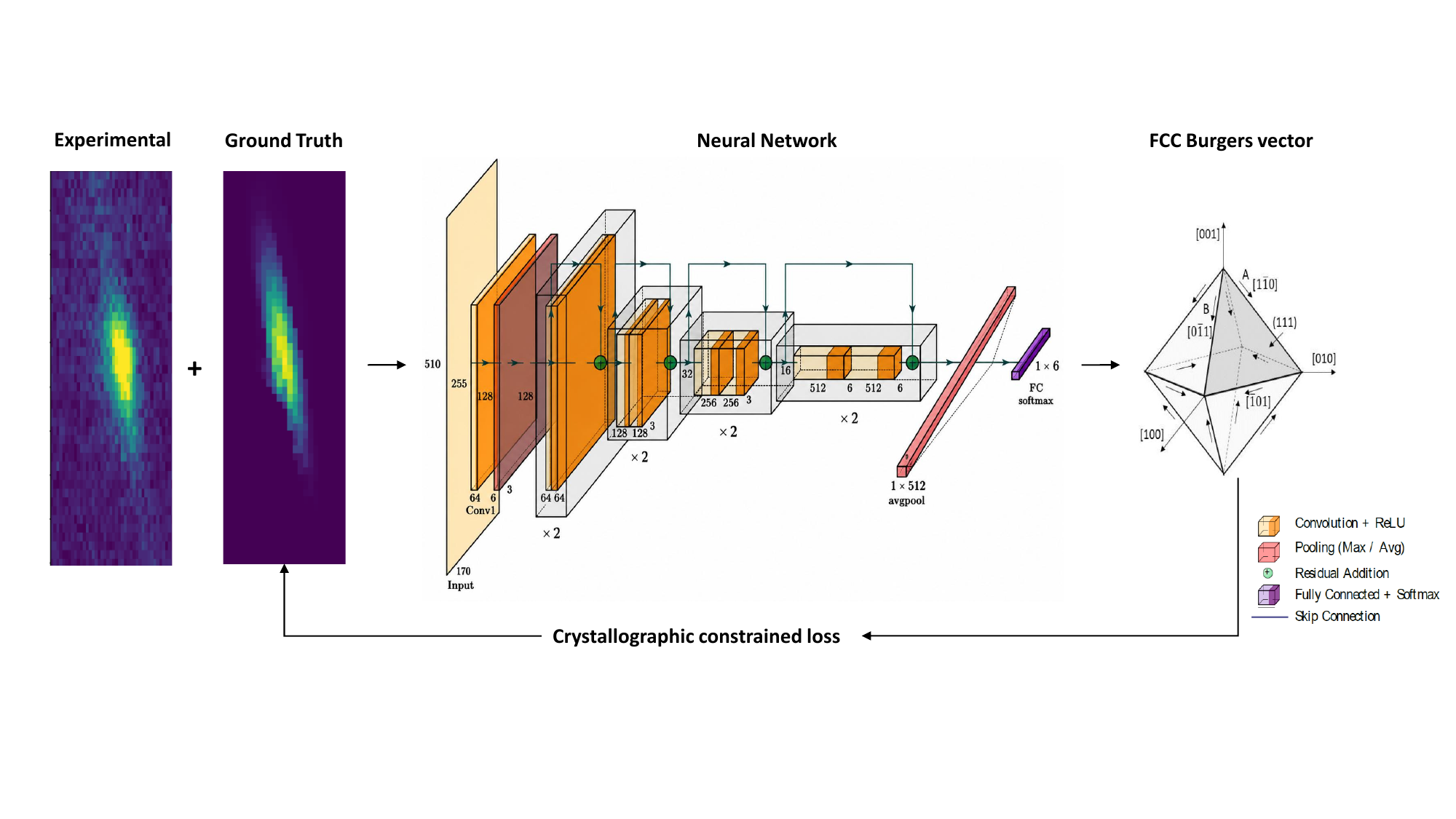}
    \caption{CNN workflow for FCC Burgers vector identification. The network is trained on simulated ground-truth WBI images using a crystallographically constrained loss, then validated on experimental dislocation images to identify the Burgers vectors.}
    \label{fig:resnet}
\end{figure}

\subsection{Convolutional Neural Network Model}\label{subsec:MachineLearningModel}

The implemented model is a CNN trained in a supervised manner. Network parameters are optimized via backpropagation to minimize a cross-entropy loss between predicted and reference Burgers-vector labels \citep{rumelhart1986,buscema1998back}. During training, batches of DFXM images are passed through the network, and outputs are compared with reference annotations. Gradients are computed and used to update the weights through stochastic gradient descent \citep{robbins1951}. This process is repeated until convergence, \textit{i.e.}, until the loss stabilizes.  

The input data consist of $510 \times 170$ pixel images generated from the GO simulations, each containing a single individual dislocation. The network architecture is based on the pre-activation ResNet-18 design (Fig.~\ref{fig:resnet}) \cite{he2016deep}. It begins with an initial convolutional layer with 16 filters, doubling at each subsequent stage up to 256. Downsampling is implemented using strided convolutions, and batch normalization with rectified linear unit (ReLU) activation is applied throughout to stabilize and accelerate training.  

After the convolutional blocks, global average pooling is applied to reduce each feature map to a single value, yielding a compact feature representation. This vector is passed through a fully connected layer that outputs scores for the six candidate Burgers-vector labels. The network is implemented in PyTorch and trained for 100 epochs with stochastic gradient descent, an initial learning rate of $1 \times 10^{-2}$ that is reduced adaptively by a scheduler, and a batch size of 16. Training took 37 min on single NVIDIA Tesla V100-SXM2-32GB graphics processing units (GPUs).

\subsection{Evaluation protocol}
The crystallographically constrained (CC) model was trained using a loss function that combines the standard data-driven
cross-entropy objective with soft crystallographic constraints. For sample
$i$, let $z_{i,c}$ denote the logit for Burgers-vector label $c$,
$p_{i,c}=\mathrm{softmax}(z_i)_c$ the corresponding predicted
probability, $\mathcal{V}_i$ the set of Burgers-vector labels
compatible with the known slip system, and $\mathcal{I}_i$ the
complementary set of incompatible labels. The loss is
\begin{equation}
\mathcal{L}
=
-\frac{1}{N}
\sum_{i=1}^{N}
\sum_{c=1}^{C}
y_{i,c}\log p_{i,c}
+
\frac{\alpha}{N}
\sum_{i=1}^{N}
\sum_{u\in\mathcal{I}_i}
p_{i,u}
+
\frac{\lambda}{N}
\sum_{i=1}^{N}
\frac{1}{|\mathcal{V}_i||\mathcal{I}_i|}
\sum_{v\in\mathcal{V}_i}
\sum_{u\in\mathcal{I}_i}
\left[
m-\left(z_{i,v}-z_{i,u}\right)
\right]_+ .
\label{eq:loss_cc}
\end{equation}

Here, $y_{i,c}$ is the one-hot target label, $[a]_+=\max(0,a)$,
$\alpha$ controls the penalty on crystallographically incompatible
probability mass, and $\lambda$ controls the logit-margin penalty. The
second term discourages probability assignment to Burgers vectors that are
incompatible with the slip system of sample $i$, while the third term
encourages compatible logits to exceed incompatible logits by a margin
$m$. These terms do not impose the crystallographic constraint exactly,
but bias the optimization toward slip-system-consistent predictions.

After training, model performance was evaluated using prediction
metrics and probabilistic uncertainty diagnostics. The predicted label was
defined as
\begin{equation}
\hat{c}_i
=
\arg\max_c p_{i,c},
\end{equation}
and the prediction accuracy was computed as
\begin{equation}
\mathrm{Accuracy}
=
\frac{1}{N}
\sum_{i=1}^{N}
\mathbb{I}\left(\hat{c}_i=c_i\right),
\end{equation}
where $c_i$ is the true Burgers-vector label and $\mathbb{I}(\cdot)$
is the indicator function. Accuracy provides a global measure of predictive
performance, but it can be insensitive to label imbalance. Therefore, the
macro-averaged F1 score was also reported. For each label $c$, let
$\mathrm{TP}_c$, $\mathrm{FP}_c$, and $\mathrm{FN}_c$ denote the number of
true positives, false positives, and false negatives, respectively:
\begin{equation}
\mathrm{Precision}_c
=
\frac{\mathrm{TP}_c}{\mathrm{TP}_c+\mathrm{FP}_c},
\qquad
\mathrm{Recall}_c
=
\frac{\mathrm{TP}_c}{\mathrm{TP}_c+\mathrm{FN}_c},
\end{equation}
and
\begin{equation}
\mathrm{F1}_c
=
2
\frac{
\mathrm{Precision}_c \, \mathrm{Recall}_c
}{
\mathrm{Precision}_c+\mathrm{Recall}_c
}.
\end{equation}
The macro-averaged F1 score is
\begin{equation}
\mathrm{Macro\ F1}
=
\frac{1}{C}
\sum_{c=1}^{C}
\mathrm{F1}_c .
\end{equation}
This metric gives equal weight to each Burgers-vector label and provides a
label-balanced measure of performance. Label-specific errors were examined using the confusion matrix $M$, where
$M_{ij}$ denotes the number of samples with true label $i$ predicted as
label $j$. This representation was used to identify systematic
substitutions between candidate Burgers vectors and to determine whether errors
were concentrated in particular labels.

The quality of the predicted probability distribution was assessed using
the negative log-likelihood (NLL),
\begin{equation}
\mathrm{NLL}
=
-\frac{1}{N}
\sum_{i=1}^{N}
\log p_{i,c_i}.
\end{equation}
Unlike accuracy, the NLL depends on the probability assigned to the true
label and penalizes confident incorrect predictions strongly. It therefore
provides a complementary measure of probabilistic performance.

Predictive uncertainty was quantified using the entropy of the predicted
label-probability distribution. For sample $i$,
\begin{equation}
H_i
=
-\sum_{c=1}^{C}
p_{i,c}\log p_{i,c}.
\end{equation}

Large entropy values indicate that the predicted probability mass is spread
over several Burgers-vector labels, whereas small values indicate more
concentrated predictions. Further, the epistemic component of uncertainty was estimated using mutual
information from $T$ stochastic forward passes. Let $p_{i,c}^{(t)}$
denote the predicted probability for label $c$ on stochastic pass $t$.
The mutual information (MI) for sample $i$ is then estimated as
\begin{equation}
MI_i
=
-\sum_{c=1}^{C}
\bar{p}_{i,c}\log \bar{p}_{i,c}
-
\frac{1}{T}
\sum_{t=1}^{T}
\left(
-\sum_{c=1}^{C}
p_{i,c}^{(t)}
\log p_{i,c}^{(t)}
\right).
\end{equation}

The mutual information measures the sensitivity of the predictive
distribution to stochastic model variations and is used here as a diagnostic
for epistemic uncertainty. Elevated values may indicate reflection
conditions or Burgers-vector labels for which the learned representation is
less certain. Together, accuracy, macro F1, the confusion matrix, NLL,
predictive entropy, and mutual information provide complementary diagnostics
of prediction performance and predictive confidence.

\begin{figure}[t]
    \centering
    \includegraphics[width=0.8\linewidth]{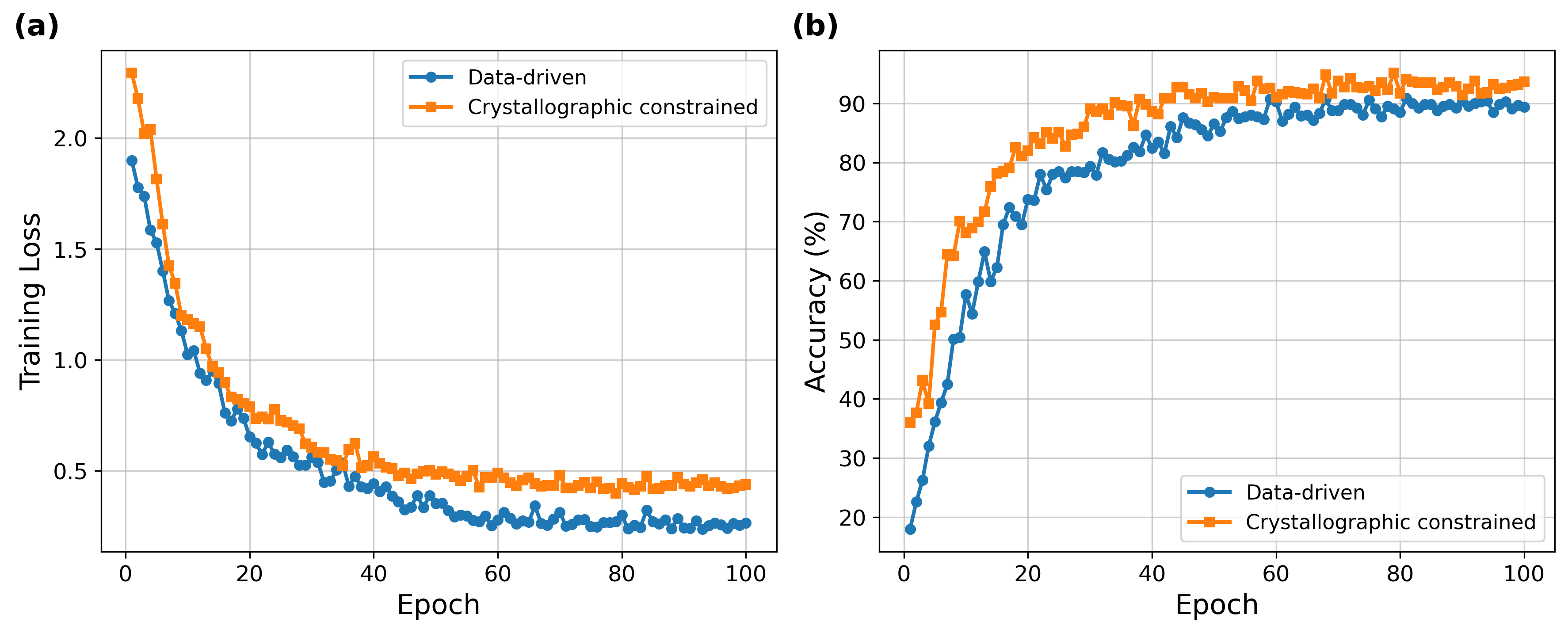}
    \caption{Training performance for the data-driven and crystallographically constrained prediction models. (a) Training loss as a function of epoch. The data-driven model reaches a lower final loss, while the constrained model stabilizes at a higher loss because the objective includes additional crystallographic penalty terms. (b) Prediction accuracy as a function of epoch. Both models converge to similar performance, with the constrained model giving slightly higher accuracy after convergence.}
    \label{fig:sim_eval}
\end{figure}

\section{Results}\label{sec:Results}

\subsection{Training and Test Performance}\label{subsec:TrainTestPerformance}

\begin{table}[ht]
\centering
\caption{Comparison of performance and uncertainty estimates for the data-driven and crystallographically constrained models. Accuracy and macro F1 quantify predictive performance, while negative log-likelihood (NLL), mean entropy, and mean mutual information (MI) summarize confidence calibration, total predictive uncertainty, and epistemic uncertainty, respectively.}
\label{tab:compact_model_comparison}
\begin{tabular}{lccccc}
\hline
\textbf{Model type} 
& \textbf{Accuracy (\%)} 
& \textbf{Macro F1 (\%)} 
& \textbf{NLL} 
& \textbf{Mean Entropy} 
& \textbf{Mean MI} \\
\hline
Data-driven 
& 88.59 
& 88.07
& 0.2556 
& 0.4000 
& 0.0660 \\
Crystallographically constrained 
& 93.53 
& 89.63 
& 0.1824 
& 0.3246 
& 0.0410 \\
\hline
\end{tabular}
\end{table}

To evaluate prediction performance, the purely data-driven model and the CC model were compared using the training curves in Fig.~\ref{fig:sim_eval} and the quantitative metrics summarized in Table~\ref{tab:compact_model_comparison}. Both models learn rapidly during the initial epochs, indicating that the dominant discriminative features are captured early in training. The data-driven model reaches a lower training loss, as expected, because it is optimized only for the image-label objective. The CC model maintains a higher loss because the objective includes additional crystallographic penalty terms. This higher loss should therefore not be interpreted directly as poorer prediction performance.

Despite its higher training loss, the CC model achieves better test performance. Its accuracy increases from 88.59\% to 93.53\%, while the macro F1 score increases from 88.07\% to 89.63\%. The larger gain in accuracy than in macro F1 indicates that the improvement is not equally distributed across all output labels, but the constrained model still gives a modest improvement in label-balanced performance. These results suggest that the crystallographic terms act as domain-informed regularization, reducing the number of physically inconsistent predictions without removing all label-specific ambiguities.

The uncertainty-related metrics follow the same trend. The negative log-likelihood decreases from 0.2556 for the data-driven model to 0.1824 for the CC model, while the mean predictive entropy and mean mutual information decrease from 0.4000 to 0.3246 and from 0.0660 to 0.0410, respectively. These values do not prove that the model is calibrated under all experimental conditions, but they show that the constraint improves both accuracy and uncertainty behavior on the simulated test set.

\begin{figure}[b!]
    \centering
    \includegraphics[width=\linewidth]{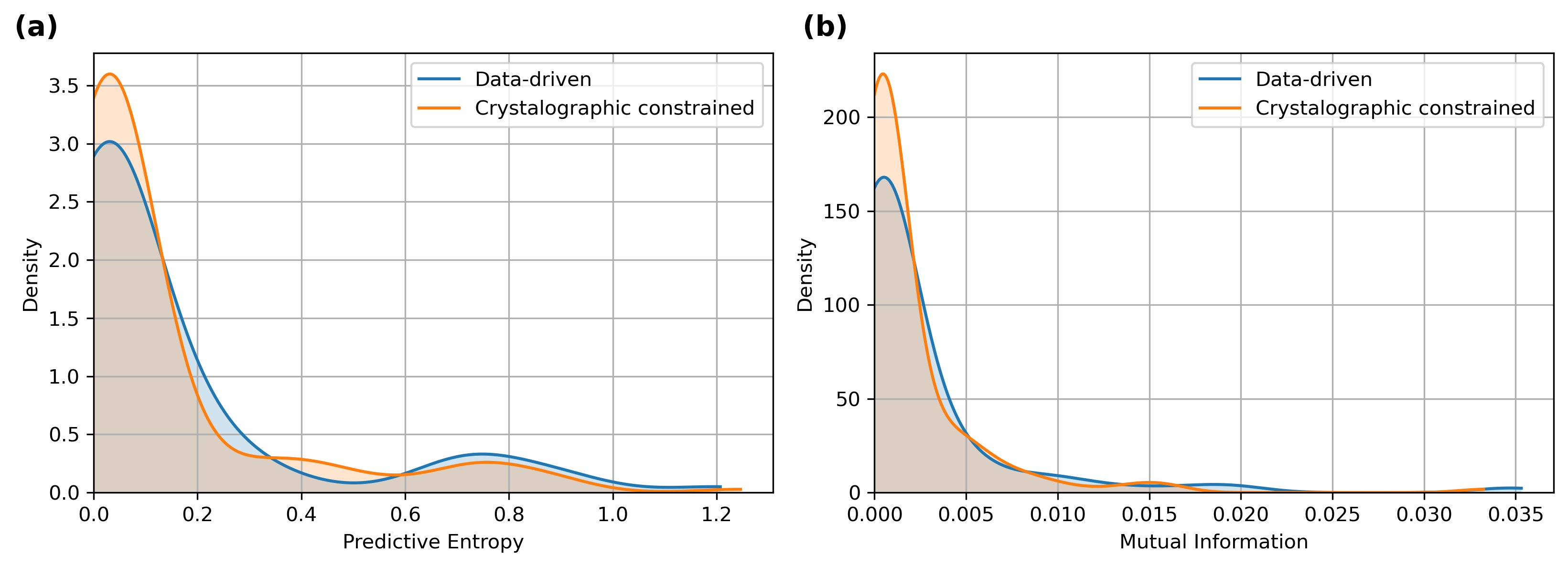}
    \caption{Uncertainty estimates for the data-driven and crystallographically constrained models on simulated test data. (a) Kernel-density estimates of predictive entropy. Both models peak at low entropy, but the data-driven model has a broader high-entropy tail. (b) Kernel-density estimates of mutual information. Both distributions are concentrated near zero, while the data-driven model shows a wider tail toward higher epistemic uncertainty.}
    \label{fig:uncertain}
\end{figure}

The predictive entropy distributions in Fig.~\ref{fig:uncertain} provide additional insight into these averages. For both models, the distributions are concentrated near low entropy values, but the data-driven model has broader high-entropy and high-mutual-information tails. The CC model reduces this spread, consistent with the interpretation that crystallographic constraints reduce ambiguous or unstable predictions.

All in all, the simulated-data results show that the CC model improves the prediction and uncertainty metrics relative to the purely data-driven model for the same WBI task. The improvement is meaningful, but it remains specific to the simulated isolated-dislocation setting and must be interpreted together with the experimental and cross-reflection tests below.

\subsection{Experimental Validation}\label{subsec:ExpValidation}

The trained CNN models were evaluated using experimental DFXM data 
from~\cite{Frankus2026}. The dataset contains three-dimensional observations 
of individual dislocations in pure aluminium, including a dislocation 
exhibiting double cross-slip. Cross-slip is a thermally activated process in 
which a screw segment transfers from its original $\{111\}$ glide plane to 
another $\{111\}$ plane containing the same Burgers vector 
\citep{Puschl2002,Lu2002,Hussein2015}. Double cross-slip involves two 
successive cross-slip events and allows a screw dislocation to bypass 
obstacles while conserving its Burgers vector.

For the experimentally observed dislocation, the two identified slip planes 
are $(1\bar{1}\bar{1})$ and $(11\bar{1})$, with respective normals
\begin{equation}
    \mathbf{n}_1=[1\bar{1}\bar{1}],
    \qquad
    \mathbf{n}_2=[11\bar{1}].
\end{equation}
Because the Burgers vector must lie in both planes, it must satisfy
\begin{equation}
    \mathbf{b}\cdot\mathbf{n}_1=0,
    \qquad
    \mathbf{b}\cdot\mathbf{n}_2=0.
\end{equation}
It must therefore be parallel to the intersection of the two planes:
\begin{equation}
    \mathbf{b}\parallel
    \mathbf{n}_1\times\mathbf{n}_2
    =
    [1\bar{1}\bar{1}]\times[11\bar{1}]
    =
    2[101].
\end{equation}
Since perfect dislocations in FCC aluminium have Burgers vectors of the type
$a/2\langle110\rangle$, the crystallographically allowed Burgers vector is
\begin{equation}
    \mathbf{b}=\pm\frac{a}{2}[101].
\end{equation}
The sign cannot be distinguished from the integrated weak-beam contrast.
This independently determined crystallographic constraint makes the 
double-cross-slip dislocation a particularly useful experimental test of the 
CNN outputs. The models were tested 
on the integrated experimental weak-beam images using the procedure described 
in Section~\ref{sec:Methods}.

\begin{figure}[t]
    \centering
    \includegraphics[width=0.7\linewidth]{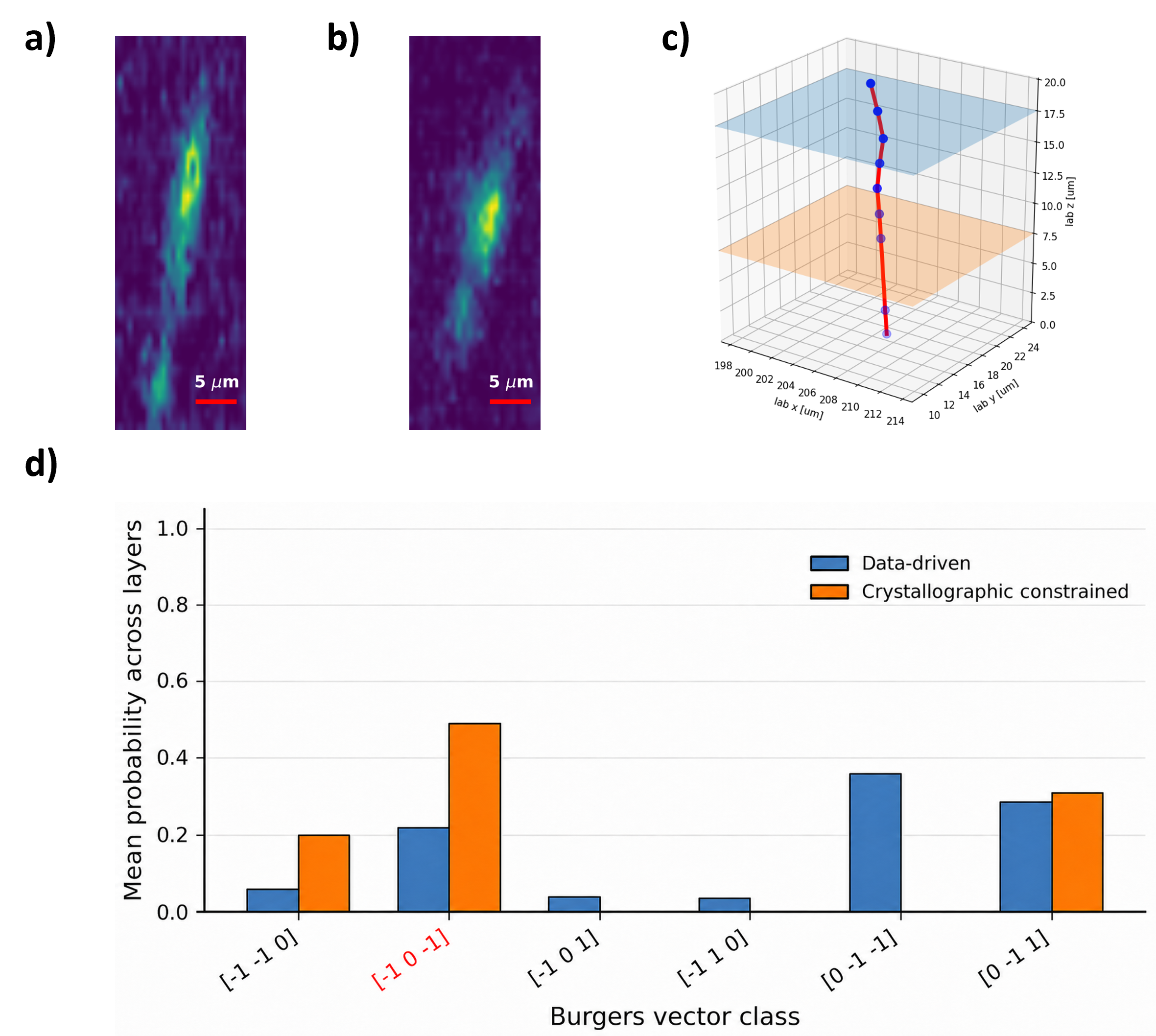}
    \caption{Experimental cross-slip case used for Burgers vector prediction. (a,b) Integrated weak-beam images from dislocation layer sections extracted from the orange and blue planes shown in (c), respectively. (c) Reconstructed 3D dislocation trajectory intersecting the two cross-slip planes. (d) Mean predicted probability across the analysed layers for the data-driven and crystallographically constrained models. The red highlighted label corresponds to the reference Burgers vector.}
    \label{fig:exp_eval}
\end{figure}

For the evaluation, regions of interest were selected manually around the dislocation signal in individual WBI images. The selected regions were cropped and preprocessed using the same procedure as for the simulated data, including resizing and intensity normalization. Figure~\ref{fig:exp_eval} shows the experimental cross-slip case used for evaluation. Figures~\ref{fig:exp_eval}a and b present WBI images from two layer sections of the same dislocation. These sections correspond to the orange and blue cross-slip planes shown in Fig.~\ref{fig:exp_eval}c. The prediction percentages for this experimental case are listed in Table~\ref{tab:cross_slip_prediction}. The reference Burgers vector of the double cross-slipping dislocation is $[\bar{1}\ 0\ \bar{1}]$. For the data-driven model, predictions are split between the reference Burgers vector, with $45.5\%$, and $[0\ \bar{1}\ 1]$, with $54.5\%$. After applying the crystallographic constraint, the reference Burgers vector becomes the dominant prediction, increasing to $72.7\%$. The remaining predictions are assigned to $[\bar{1}\ \bar{1}\ 0]$ and $[0\ \bar{1}\ 1]$, with $9.1\%$ and $18.2\%$, respectively. The constraint therefore shifts the experimental prediction distribution toward the Burgers vector expected from the cross-slip geometry.

\begin{table}[ht]
\centering
\caption{
Prediction performance for the experimental cross-slip case across 10 layers. The reference Burgers vector is $[\bar{1}\ 0\ \bar{1}]$.
}
\label{tab:cross_slip_prediction}
\begin{tabular}{llc}
\hline
Model & Predicted Burgers vector & Prediction percentage (\%) \\
\hline
Data-driven 
& \textbf{$[\bar{1}\ 0\ \bar{1}]$} & \textbf{45.5} \\
& $[0\ \bar{1}\ 1]$ & 54.5 \\
\hline
Crystallographically constrained 
& $[\bar{1}\ \bar{1}\ 0]$ & 9.1 \\
& \textbf{$[\bar{1}\ 0\ \bar{1}]$} & \textbf{72.7} \\
& $[0\ \bar{1}\ 1]$ & 18.2 \\
\hline
\end{tabular}
\end{table}

\section{Discussion}

The results show that incorporating crystallographic information into the learning process improves Burgers vector identification from WBI DFXM images of isolated dislocations. The CC model optimizes the image-based prediction task while also penalizing candidate Burgers vectors that are incompatible with the known slip geometry. Its improved test performance and uncertainty metrics indicate that this constraint acts as a useful form of domain-informed regularization for the simulated test set. In practice, uncertainty metrics are useful because they can flag predictions that should be treated with caution or selected for further inspection. However, the uncertainty analysis here is based on the simulated test distribution and Monte Carlo dropout; additional calibration on larger experimental datasets is needed before these values can be used as absolute confidence estimates.

The experimental double cross-slip case provides a first test of transfer from simulated weak-beam images to real DFXM data. This is a demanding case because the dislocation changes slip plane and line direction while retaining the same Burgers vector. The constrained model assigns the largest fraction of predictions to the reference Burgers vector, supporting the usefulness of the crystallographic constraint. The result should nevertheless be interpreted as a validation case rather than a broad experimental benchmark. The remaining disagreement with the reference Burgers vector is likely related to the simulation--experiment gap, including noise, background variations, intensity overlap with neighbouring dislocations, and changes in dislocation contrast as the dislocation character varies along the line.

The reflection-specific results in Fig.~\ref{fig:sim_cross_eval} provide insight into what the CNN learns from the training data. Although models trained and tested on the same reflection perform well, direct transfer between the FCC $(002)$ and $(111)$ reflections is poor. This behaviour is consistent with the reflection-dependent invisibility criterion: for a given diffraction vector, Burgers vectors satisfying $\nvec{Q}\cdot\nvec{b}=0$ produce little or no dislocation contrast and are therefore absent or weakly represented in the corresponding training data. The cross-reflection confusion matrices suggest that the CNN learns both the morphology of visible dislocations and this reflection-specific visibility rule. When applied to a different reflection, it continues to suppress Burgers vectors that were invisible under the training geometry, even when they are visible under the test geometry. Thus, the poor transfer reflects an internalized $\nvec{Q}\cdot\nvec{b}$ prior rather than purely random uncertainty or ambiguous contrast.

This result demonstrates that the network extracts physically meaningful information from the simulated DFXM images, but also reveals a visibility-dependent bias. Reliable inference therefore requires training data that represent the experimental diffraction geometry, detector response, and weak-beam integration procedure. The crystal orientation must also be known a priori because different azimuthal angles can produce different dislocation contrasts for the same Burgers vector, as illustrated in Fig.~\ref{fig:topotomo}. The ESRF beamline \citep{Isern2025} is particularly suitable for this approach because it enables 3DXRD orientation mapping to be combined with DFXM imaging \citep{shukla2025}, providing the required orientation information. Future models could incorporate the diffraction vector, reflection index, and sample orientation as explicit inputs. Reflection-aware or multi-branch architectures \citep{ngiam2011} may further help distinguish Burgers-vector information from reflection-dependent visibility.

Finally, the present study is restricted to isolated dislocations in FCC aluminium and to WBI rocking curves for which geometrical-optics simulations provide a suitable training basis. Extension to crowded dislocation structures and overlapping contrast fields will require moving from cropped single-dislocation classification toward joint detection and identification. Overall, the results establish the feasibility of simulation-trained, crystallographically constrained CNNs for Burgers-vector identification while defining the experimental conditions and methodological developments required for broader application.

\section{Conclusion}

We have demonstrated CNN-based Burgers vector identification of isolated dislocations in WBI DFXM images using models trained entirely on GO simulations. Introducing crystallographic constraints improves simulated test performance and reduces uncertainty estimates, showing that physically informed learning can regularize the model beyond a purely data-driven image-label objective. The experimental cross-slip case shows that the constrained model can transfer to real DFXM data in a representative isolated-dislocation example, even when the local slip geometry changes along the dislocation line. Together, these results define a practical path toward simulation-trained and uncertainty-aware Burgers vector identification, while emphasizing the need for reflection-aware models and broader experimental validation.


\section*{Acknowledgments}
We acknowledge support by the European Union's Horizon 2020 Research and Innovation Programme under the Marie Sklodowska-Curie COFUND scheme (grant agreement No. 101034267), by the European Research Council (ERC) Advanced Grant No. 885022, and by the European Spallation Source (ESS) Lighthouse on Hard Materials in 3D, SOLID (Danish Agency of Science and Higher Education, grant No. 8144-00002B). We also thank the ESRF for granting the beamtime (https://doi.org/10.15151/ESRF-ES-1005347817) on ID06-HXM for the experiment.

\section*{Data availability}

The code used to generate the ground-truth data is available at
\url{https://github.com/borgis/Geometrical_Optics_master}.
The code used to estimate the Burgers vector is available at
\url{https://github.com/Benhadjira/dfxm-bv}.

\appendix

\section{Supplementary materials}\label{app:supplementary}

\begin{figure}
    \centering
    \includegraphics[width=\linewidth]{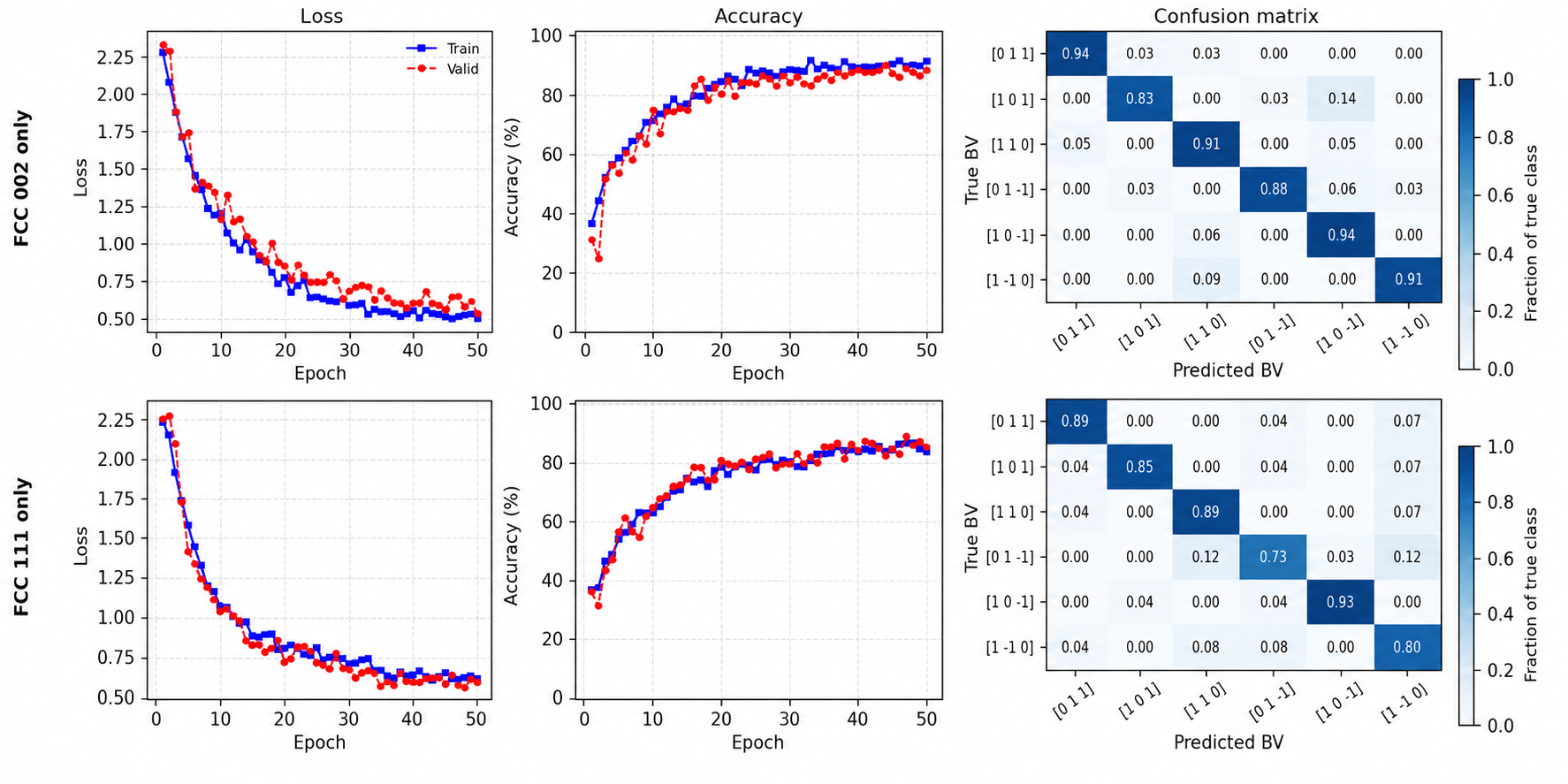}
    \caption{Single-reflection training performance on simulated WBI images. The top row shows a model trained and evaluated using only the FCC $(002)$ reflection, and the bottom row shows the corresponding result for the FCC $(111)$ reflection. For each reflection, the left and middle panels show training and validation loss and accuracy over 50 epochs, respectively. The right panels show normalized confusion matrices for the six Burgers-vector labels. Both single-reflection models learn the prediction task, but the label-wise confusion patterns differ between reflections, showing that the available contrast and label separability depend on diffraction geometry.}
    \label{fig:single_ref_eval}
\end{figure}

The supplementary figures provide additional information on reflection-specific training and transfer between diffraction reflections. Figure~\ref{fig:single_ref_eval} shows the training loss, validation accuracy, and normalized confusion matrices for models trained and evaluated on a single reflection. The $(002)$-only model reaches higher label-wise agreement than the $(111)$-only model, while the $(111)$-only model shows stronger off-diagonal confusion for several candidate Burgers vectors.

\begin{figure}
    \centering
    \includegraphics[width=0.7\linewidth]{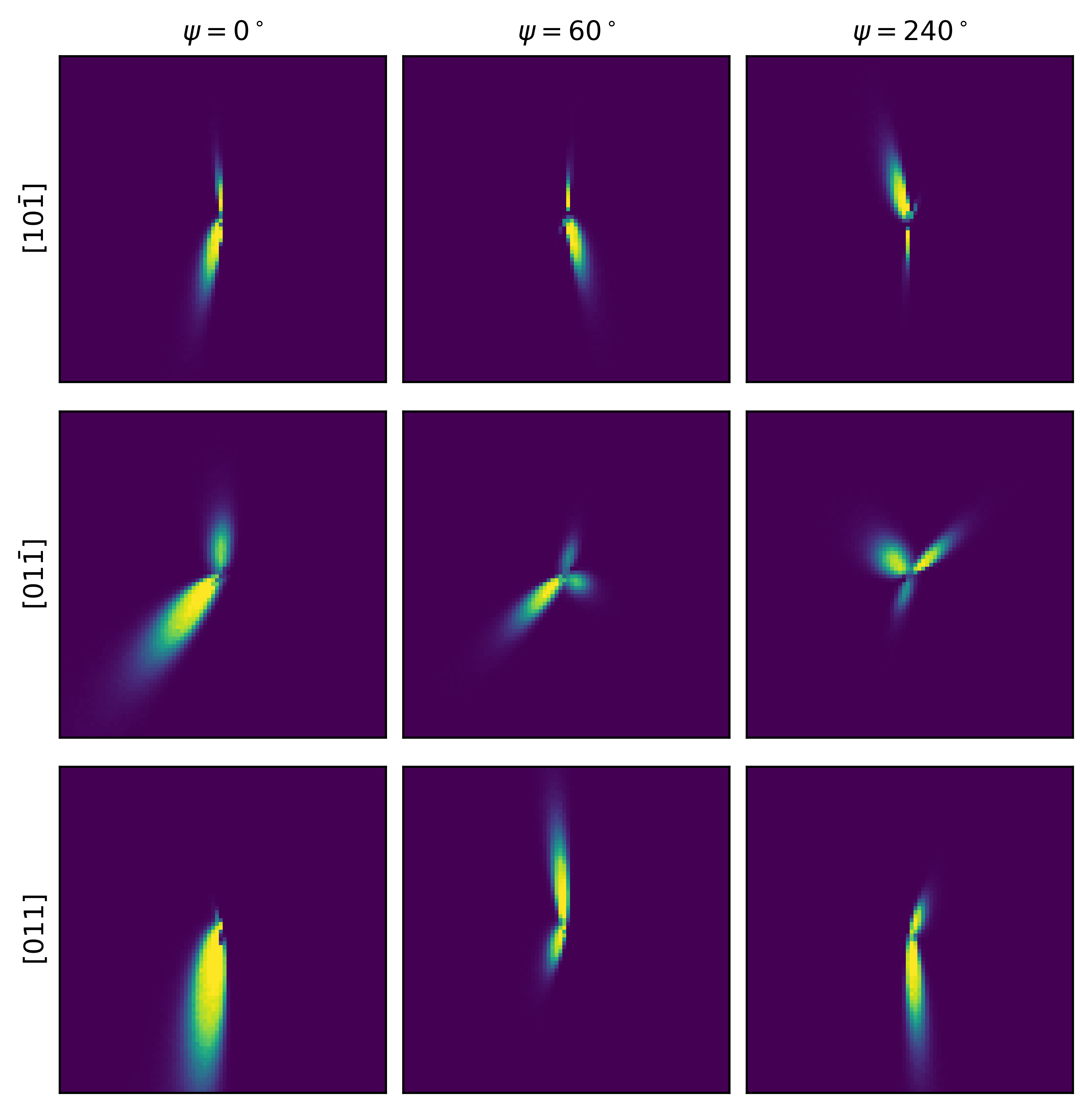}
    \caption{Simulated DFXM contrast for three FCC $a/2\langle110\rangle$ dislocations measured at the fixed $\mathbf{q}=111$ reflection. Rows correspond to Burgers-vector directions $[10\bar{1}]$, $[01\bar{1}]$, and $011$, while columns show crystal rotations of $\psi=0^\circ$, $60^\circ$, and $240^\circ$. The variation across rows and columns reflects the dependence of the DFXM contrast on both Burgers-vector orientation and the azimuth rotation angle}
    \label{fig:topotomo}
\end{figure}

Figure~\ref{fig:topotomo} illustrates the dependence of the simulated DFXM contrast on the azimuthal angle $\psi$, defined here as rotation about the fixed diffraction vector. For a given Burgers vector, changes in $\psi$ can produce substantial changes in the projected contrast morphology and intensity distribution. Hence, models should be trained using simulations generated at the appropriate experimental azimuthal angle.

\begin{figure}
    \centering
    \includegraphics[width=\linewidth]{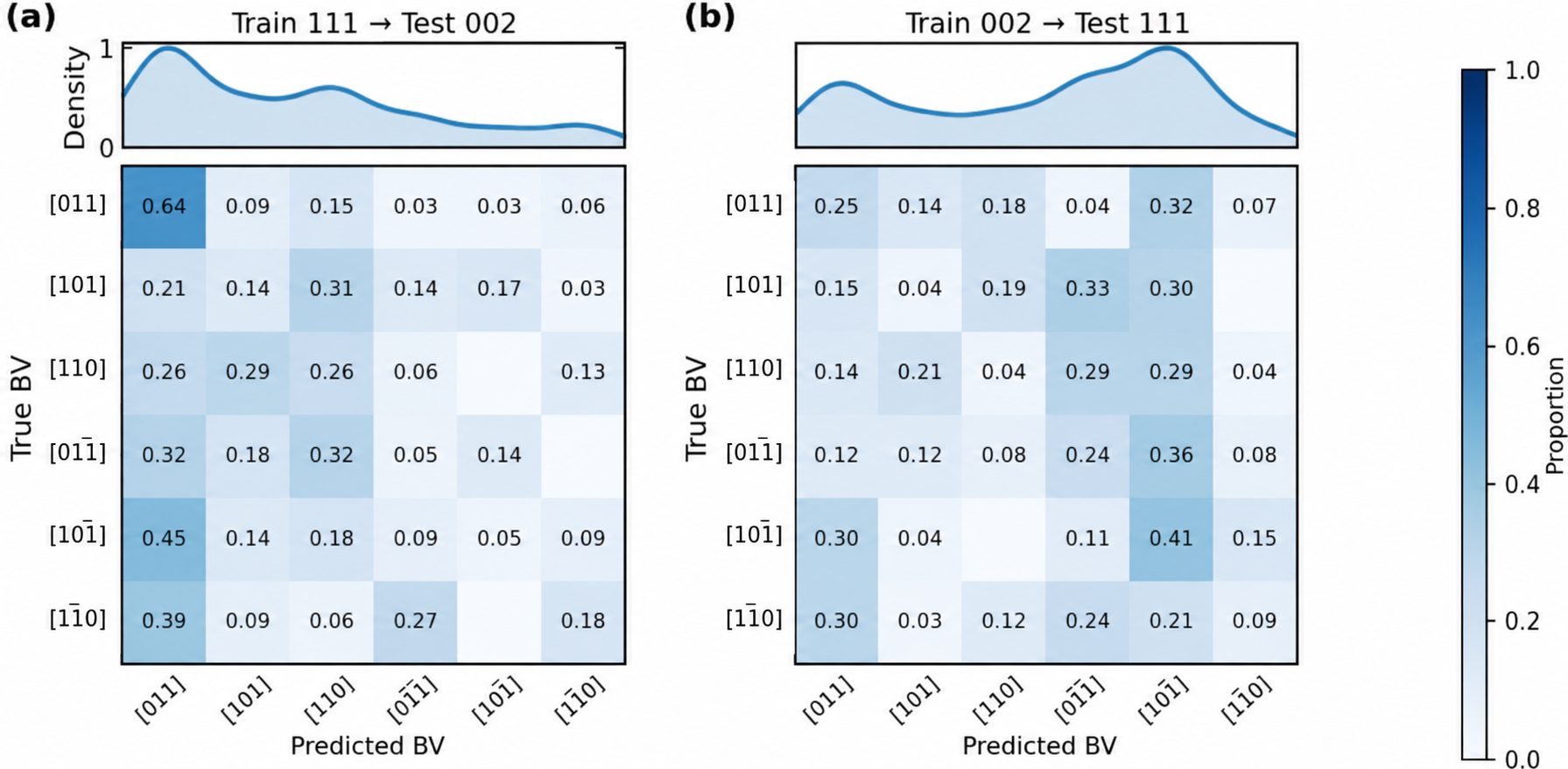}
    \caption{Cross-reflection confusion matrices on simulated data. (a) Model trained on the FCC $(111)$ reflection and tested on the FCC $(002)$ reflection, giving an accuracy of 17.6\%. (b) Model trained on the FCC $(002)$ reflection and tested on the FCC $(111)$ reflection, giving an accuracy of 15.7\%. The prediction-density distributions above the matrices show that the models retain reflection-specific label preferences when transferred to the other reflection, consistent with the reflection dependence of the $\nvec{Q}\cdot\nvec{b}$ visibility condition.}
    \label{fig:sim_cross_eval}
\end{figure}

Figure~\ref{fig:sim_cross_eval} tests whether a model trained on one reflection can be applied directly to a different reflection. The low cross-reflection accuracies show that this transfer is not reliable for the present training setup and motivate future models that include reflection information explicitly.

\bibliographystyle{elsarticle-harv} 
\bibliography{references}

\end{document}